\documentclass[12pt]{iopart}
\usepackage{graphicx}

\begin{document}

\title{Collision Assisted Zeeman Cooling with Multiple Types of Atoms}
\author{R C Ferrier, M S Hamilton and J L Roberts}
\address{Department of Physics, Colorado State University, Fort Collins, Colorado 80523, USA}
\ead{rebekahferrier@gmail.com}

\begin{abstract}
Through a combination of spin-exchange collisions in a magnetic field and optical pumping, it is possible to cool a gas of atoms without requiring the loss of atoms from the gas.  We investigated this technique, Collision Assisted Zeeman Cooling, using $^{85}$Rb and $^{87}$Rb. We experimentally confirmed that our measured two isotope CAZ cooling rate agreed with a cooling rate predicted though a simple analytic model. As part of the measured cooling rate, we quantitatively characterized the heating rates associated with our actual implementation of this cooling technique and found hyperfine-changing collisions to be a significant limitation for the $^{85/87}$Rb gas mixture. We comment on the prospects for improving the cooling performance beyond that presented in this work.
\end{abstract}


\submitto{\jpb}
\maketitle


There are numerous techniques ~\cite{davis:1995,masuhara:1988,price:2008,ferrari:2001,fattori:2006,medley:2011,moler:1992,kerman:2000,perrin:1999,barrett:2001,holland:2000,granade:2002,monroe:1995,wolf:2000,aspect:1988,boyer:2004,petsas:1996,stecher:1997,ivanov:2012,dunn:2005,C-T:2011}that have been theoretically and/or experimentally evaluated for cooling gasses of ultracold atoms below the limits associated with standard laser cooling techniques such as Doppler cooling ~\cite{hansch:1975,wineland:1979,chu:1985} and optical molasses cooling ~\cite{lett:1988,dalibard:1989}. By far the most common of these cooling techniques is evaporative cooling and as such it has a long history of being effective, robust, and straightforward to implement ~\cite{davis:1995,masuhara:1988}. There are reasons to investigate non-evaporative cooling techniques, however. There is intrinsic physics interest in studying the ways in which ultracold gases can be cooled. Evaporative cooling requires the loss of atoms from the trapped gas, which is generally undesirable. A rapid non-evaporative cooling scheme would have the potential to simplify experimental apparatuses by eliminating, for instance, the need for atom transport between different regions of the vacuum system~\cite{Lewandowski}. Finally, in non-evaporative cooling the link between the cooling mechanism and the potential confining the ultracold atoms is reduced, allowing for more flexibility in tailoring the confining potential~\cite{barrett:2001}.

In previously published work, a non-evaporation-based cooling technique called Collision-assisted Zeeman (CAZ) cooling was presented theoretically~\cite{ferrari:2001}. At the heart of this cooling technique are spin-exchange collisions that transfer kinetic energy to Zeeman energy that is then subsequently removed from the ultracold gas via optical pumping. We present an extension to CAZ cooling by considering CAZ cooling with two different types of atoms present in the gas being cooled rather than one. We refer to this two-type-of-atom based cooling scheme as 2-CAZ. While cooling two different types of atoms introduces complexity, it leads to predicted advantages for CAZ cooling both specific to the cooling technique itself and also in more general terms through a predicted improvement in cooling rate at low temperatures. In addition to presenting 2-CAZ theoretically, the experimental implementation of 2-CAZ in an $^{85}$Rb/$^{87}$Rb system is reported. While optical trap loading behaviour and hyperfine-changing collisions prevented robust 2-CAZ cooling in our system, cooling rate measurements were made that allow for a comparison between experimentally realizable and theoretically predicted 2-CAZ performance. Through these measurements, requirements for robust 2-CAZ cooling are better characterized and understood.

\section{2-CAZ cooling and 2-CAZ cooling rate}

Before describing 2-CAZ cooling, it is useful to briefly summarize CAZ cooling as presented in Ref~\cite{ferrari:2001}. Figure 1 shows the relevant states needed to describe CAZ cooling for an atom with a nonzero ground state hyperfine interaction that has electron angular momentum $L$=0, an electron spin $S$=1/2, and nuclear spin $I$=3/2 in its $F=L+S+I=1$ ground state. An external magnetic field is applied that causes a Zeeman energy shift for the different angular momentum projections (quantum number $m_F$) along the magnetic field direction. Because of \textit{second-order} Zeeman shifts, the Zeeman energy of the $m_F$=0 state is less than the average of the Zeeman energies of the $m_F$=1 and $m_F=-1$ states.  Cooling occurs through optically pumping all of the atoms in the gas into the $m_F$=0 state. These atoms then collide via spin-exchange collisions~\cite{happer:1972}. These collisions arise due to exchange effects leading to different collision phase shifts between triplet and singlet electron spin states of the colliding atom pair and result in a change of the individual $m_F$ states of a colliding atom pair while preserving the sum of the $m_F$ quantum numbers of the colliding pair ~\cite{Stoof1988}. Thus, for the states in fig. 1, two $m_F$=0 atoms can undergo a spin-exchange collision to exit the collision as an $m_F$=1 atom and an $m_F=-1$ atom. The resulting increase in Zeeman energy comes through the reduction of the kinetic energy of the colliding pair. Upon being optically pumped back to the $m_F$=0 state, these two atoms are returned to their starting states, but with less kinetic energy and so cooling results. Heat is carried away by the photons scattered during the optical pumping process and no atom loss is required as part of the cooling cycle. This type of cooling has similarities to demagnetization cooling in ultracold gases~\cite{fattori:2006} and nuclear demagnetization cooling in solid state systems~\cite{kurti}.

\begin{figure}
\includegraphics{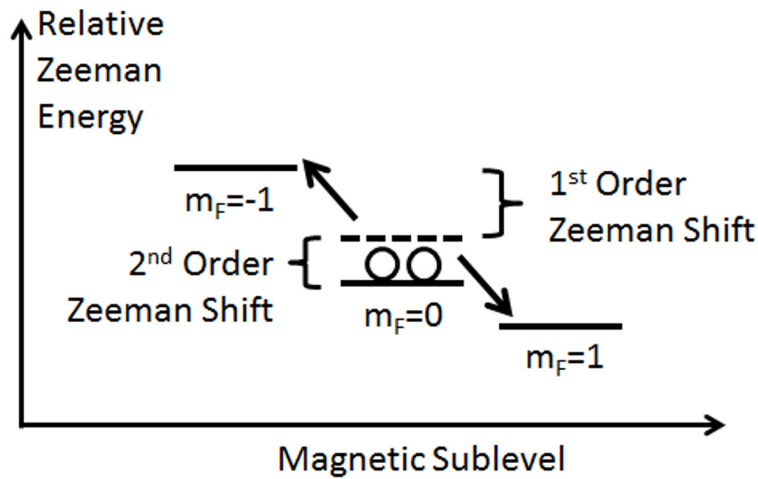}
\caption{Relevant states for CAZ cooling as presented in Ref~\cite{ferrari:2001} for reference in describing spin-exchange cooling with a single type of atom as a precursor to two-type-of-atom cooling that is the subject of this work. The solid lines depict Zeeman shifted $m_F$ states while the dotted line indicates the average energy of the $m_F$=+1 and $-1$ states. The circles are meant to represent the states of two atoms at the start of the cooling cycle and the arrows indicate the change in state due to spin-exchange collisions.}
\end{figure}

Extending this cooling scheme to two different types of atoms in 2-CAZ is straightforward. The general idea is the same, except that now the spin-exchange collisions occur between the two different types of atoms. For concreteness, consider 2-CAZ with atoms like $^{85}$Rb ($I=5/2$) and $^{87}$Rb ($I=3/2$) in their lower hyperfine states ($F$=2 and 1, respectively). Although, it is important to note that unlike CAZ cooling it is not a requirement that the atoms in 2-CAZ have hyperfine structure, just non-zero total electron angular momentum. Figure 2 shows a 2-CAZ cooling cycle starting with an $^{85}$Rb atom in the $F$=2, $m_F=-2$ state and an $^{87}$Rb atom in the $F=1$, $m_F$ =0 state. Spin-exchange collisions will scatter these atoms into their $m_F=-1$ states. Because of the associated $g_F$ factors for $^{85}$Rb and $^{87}$Rb, the total Zeeman energy increases by $\frac{1}{6}\mu_B B$ as a result of this collision. ($\mu_B$ is the Bohr magneton, $B$ is the externally applied magnetic field). Just as before, the increase in Zeeman energy is supplied via the kinetic energy of the collision. To close the cooling cycle, the $^{85}$Rb atoms are optically pumped back to the $m_F=-2$ state. For $^{87}$Rb, RF fields are used to drive transitions between $^{87}$Rb $m_F$ states to scramble their populations periodically. Spin-exchange collisions also occur between the $^{85}$Rb $F$=2,$m_F=-2$ atoms and $^{87}$Rb F=1, $m_F$=1 atoms that change the $|m_F|$ of the individual atoms by both 1 and 2, but for optimal cooling parameters the $\Delta|m_F|=1$ collisions are dominant.

For both CAZ and 2-CAZ cooling, the advantages are that the spin-exchange collisions occur spontaneously, they are naturally three-dimensional, the main cooling adjustment is a magnetic field, the energy removed per scattered optical pumping photon is proportional to the gas temperature, and relatively weak optical pumping is sufficient for cooling. The main disadvantage is that a sufficient atom density is needed for a fast enough cooling rate; and that turns out to be a serious limitation in our experimental realization of this technique. Also, reabsorption effects have the potential to limit the achievable cooling rate at lower temperatures, as with other light-based sub-Doppler cooling schemes ~\cite{Castin1998,Cirac1996,Wolf2000} (although this is less of a problem with 2-CAZ as described below). Finally, the use of spin-exchange collisions means that $m_F$-dependent traps such as magnetic traps cannot be used with this cooling. Instead $m_F$-insensitive traps like optical traps are required.

The main disadvantage of 2-CAZ vs. CAZ cooling is the experimental complexity of using two different types of atoms. There are several offsetting advantages, however. In 2-CAZ, the first-order rather than the second-order Zeeman effect can be used, resulting in smaller magnetic fields being required for optimal cooling. Optimal cooling occurs when the change in Zeeman energy in the spin-exchange collision($\Delta$) is roughly equal to $k_B T $, where $T$ is the gas temperature, assuming negligible heating owing to the the optical pumping process. For the specific 2-CAZ cooling configuration shown in fig. 2, for $\Delta$=$k_B$30$\mu K$ a field of 2.7 $G$ is required. In contrast, if $^{85}$Rb were to be CAZ cooled alone, a field of 66 $G$ would be required for the same $\Delta$. Not only is the lower field easier to produce experimentally, but at 66 $G$ the optical pumping required to close the CAZ cooling loop is complicated by the fact that Zeeman shifts are on the order of the excited state hyperfine energy splittings. Also, at higher magnetic fields the probability of undesirable heating collisions, such as dipole relaxation ~\cite{Stoof1988}, increases.

\begin{figure}
\includegraphics{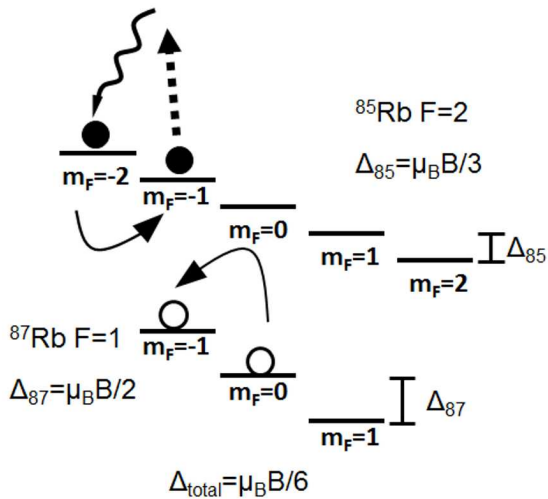}
\caption{Collision-Assisted Zeeman Cooling of $^{85}$Rb/$^{87}$Rb. This diagram depicts 2-CAZ cooling in an $^{85/87}$Rb mixture.  Spin-exchange collisions change $^{85}$Rb $F=2,m_F=-2$+$^{87}$Rb $F=1,m_F=0$ ($F=1,m_F=1)$ atom pairs to $^{85}$Rb $F=2,m_F=-1$+$^{87}$Rb $F=1,m_F=-1$ ($F=1,m_F=0)$ pairs. Each collision results in $\frac{1}{6}\mu_B B$ of kinetic energy being removed from the sample. The $^{85}$Rb is optically pumped back into its initial state after a collision, keeping it spin-polarized.  This ensures that only cooling collisions take place. The Zeeman shift coefficient for $^{85}$Rb and $^{87}$Rb are labeled as $\Delta_{85}$ and $\Delta_{87}$ respectively.}
\end{figure}

There are other advantages of 2-CAZ cooling vs. just CAZ cooling besides being able to use a smaller magnetic field. The use of two different types of atoms allows for a search for a pair with particularly favorable spin-exchange collision rates. More importantly, the fact that optical pumping needs only be performed on one of the two species in the gas has multiple advantages. 2-CAZ cooling can be performed even in the situation where one of the species is not amenable to optical pumping (e.g. with molecules). The optically-pumped species can be kept optically thin during the cooling through adjusting its number. In the presence of optical-density-dependent heating through reabsorption during the optical pumping part of the cooling cycle, the ability to reduce the number of atoms in the optically-pumped species is expected to extend effective cooling to lower temperatures.

Despite the number of $m_F$ states in a system like that in fig. 2, it is possible to write a simple analytical expression for the 2-CAZ cooling rate that has a wide range of validity. This is possible through using detailed balance considerations and the fact that certain collision channels are dominant for the magnitude of magnetic fields likely to be used. In this analytic expression it is assumed that the two different atoms are in sufficient thermal contact so that their temperatures are not widely different, that both gases are in kinetic thermal equilibrium, that the applied magnetic field is not too far from the optimal value, that there is not a strong (i.e. much greater than factor of 2) variation in spin-exchange rates as a function of $m_F$ state, and that the non-optically pumped atoms' $m_F$ populations are rebalanced frequently enough that they do not deviate significantly from their average values during the cooling. All of these assumptions are reasonable in the $^{85/87}$Rb mixture we investigated experimentally and can be expected to apply to many other gas mixtures as well. In addition, it is assumed that the atoms are confined in a harmonic potential. The instantaneous 2-CAZ cooling rate under these assumptions is

\begin{equation}
\frac{dT}{dt}=-\frac{1}{\tau_{SE}}\frac{exp(-\frac{\Delta}{k_B T})N_2}{3k_B(N_1+N_2)}(\Delta - \kappa)\frac{1}{1+\tau_{OP}/\tau_{SE}(1+exp(-\frac{\Delta}{k_B T}))}
\end{equation}

\noindent where again $\Delta$ is the change in Zeeman energy in a spin-exchange collision that changes $|m_F|$ by 1 ($\mu_B B/6$ for $^{85/87}$Rb), $T$ is the temperature of the gas, and $k_B$ is Boltzmann's constant. $N_1$ and $N_2$ are the atom numbers of the non-optically-pumped and optically-pumped atoms, respectively. The spin-exchange time-constant is defined such that $\frac{1}{\tau_{SE}}=k_2 n_1$ where $n_1$ is the average density of the non-optically-pumped atoms and $k_2$ is the spin-exchange collision rate weighted assuming equal $m_F$ populations for the non-optically-pumped atoms. $\tau_{OP}$ is the 1/$e$ time associated with the optical pumping mechanism. Finally, $\kappa$ is the average of the energy that is imparted per \textit{successful} optical-pumping-driven population transfer.

Heating must occur during optical pumping since photons are spontaneously scattered. Each photon scatter results in a random recoil momentum kick being imparted to the scattering atom that on average increases its kinetic energy. In addition, $\kappa$ can also depend on the densities of the atoms in the gas. For instance, if the gas of optically-pumped atoms ($N_2$) is optically thick, then reabsorption will increase the amount of energy imparted per optical pumping cycle as photons scatter multiple times before leaving the gas ~\cite{Castin1998,Cirac1996,Wolf2000}. Additionally, density-dependent collisions may produce not only losses but heating during the optical pumping cycle. These heating mechanisms represent a limit on the lowest achievable temperatures.

Since the value of $\Delta$ is set by the strength of the applied magnetic field, for any set of conditions the cooling rate can be maximized. In the limit of fast optical pumping (i.e. $\tau_{OP}$ goes to zero), the optimal value of $\Delta=k_BT+\kappa$ and the optimal cooling rate is

\begin{equation}
\frac{dT_{opt}}{dt}=-\frac{1}{\tau_{SE}}\frac{exp(-1-\frac{\kappa}{k_B T})N_2}{3(N_1+N_2)}T
\end{equation}

\noindent  We note that the cooling rate varies little in a fractional sense around the optimal value of $\Delta$, and so setting $\Delta$ to precisely its optimal value is not a critical requirement for effective cooling.

\section{Measurement of 2-CAZ cooling rate in an ${85/87}$Rb mixture}

If equation (1) is an adequate description of the 2-CAZ cooling rate, then it can be used to evaluate the effectiveness of 2-CAZ cooling for a variety of experimental configurations. We implemented 2-CAZ cooling in a gas composed of both $^{85}$Rb and $^{87}$Rb simultaneously loaded into a Far Off-Resonance Optical Trap (FORT)  to measure the cooling rate that could be obtained using this technique. Using two isotopes of Rb had the advantage that similar laser systems and optics were used in the initial laser cooling of the atoms. Also, the predicted spin-exchange rate between these isotopes is favorable. Our experimental techniques and cooling rate measurements are reported in this section.
  
The experimental sequence started with loading $^{85}$Rb and $^{87}$Rb atoms simultaneously into overlapping magneto-optical traps (MOTs). From the MOTs, they are then loaded into the FORT~\cite{Miller1993}. Once in the FORT, we turn on a uniform magnetic field of 2 G. From this point, we made several different types of measurements and conducted various experiments, including: characterizing the optical pumping efficiency, measuring heating and loss during optical pumping, measuring evaporation rates, measuring the initial $m_F$ state populations after initial spin polarization, measuring the 2-CAZ cooling rate through monitoring the $^{85}$Rb F=2, $m_F=-1$ state population, and examining collisional loss rates in the gas. All of these measurements used one or more experimental capabilities, described below, that we built into our apparatus.  

\textbf{MOT/FORT.}
The $^{85}$Rb and $^{87}$Rb MOTs were created using standard techniques~\cite{Raab1987}. The atoms were loaded from the MOTs into the FORT using a compressed MOT stage by reducing the hyperfine repump power and detuning the main trapping laser further to the red of the cycling transition~\cite{Petrich1994}. The FORT is created using an AOM-controlled 75W 10.64$\mu$m CO$_2$ laser focused to a spot that results in a FORT with a trap depth of 280$\mu$K  and trap frequencies of 855Hz in the radial direction and 30Hz in the axial direction. The radial frequency was measured using a parametric heating technique~\cite{Wu2006,Roati2001,Friebel1998}.

To extract number and temperature information from the atoms, the FORT was turned off rapidly and atoms were allowed to expand for 3-5ms. The cloud was then imaged onto a CCD camera using absorptive imaging on the atoms' cycling transition. Normally we applied hyperfine repump light during the imaging to measure the total atom number.  For some measurements, though, we imaged without hyperfine repump light to measure just the atoms in their upper hyperfine state. With our system, we could only measure the number and temperature of one of the two Rb isotopes per experimental sequence. 

\textbf{AH Coils/2G Field.}
The Anti-Helmholtz coils used to create the MOTs were reused to create a uniform magnetic field by reversing the current direction in one of the coils using a set of mechanical relay switches. Most of our data were collected at a field of 2 $G$ because it was convenient for our apparatus and because it was close to the optimum cooling value for much of our cooling measurements. We also performed measurements at 1 $G$ to confirm our 2 $G$ results.

\textbf{Microwave measurement of $^{85}$Rb $m_F$ state distribution.}
A microwave signal was created using a microwave generator and amplifier that is connected to a microwave antenna inside of the vacuum chamber. The microwave frequency could be tuned to any of the individual $^{85}$Rb ground state F=2 to F=3 $\Delta$m$_F$=0 transitions. All of these transition frequencies are non-degenerate. 

\textbf{Spin Polarization Techniques.}
2-CAZ cooling in our system required optical pumping of $^{85}$Rb in its F=2 state from $m_F=-1$ to $m_F=-2$.
Since this could not be done through closed transitions, any useful method needed to address $^{85}$Rb atoms in both the F=2 and F=3 states. We investigated two different methods to spin polarize the $^{85}$Rb atoms. The first was an all-optical pumping method, but we found it created a large amount of heat and loss. We then tried a combination of microwave sweeps and optical pumping and found that the losses were far lower than the all-optical case.

\textbf{All Optical Pumping.}
The simplest implementation of an all optical pumping scheme used two lasers (figure 5a) . The first, L1, was pulsed and was detuned 12MHz to the blue of the $^{85}$Rb 5S$_{1/2}$ F=2 to 5P$_{1/2}$ F'=2 transition (i.e. near the D1 line). The second laser, L2, was detuned 24MHz to the blue of the 5S$_{1/2}$ F=3 to 5P$_{3/2}$ F’=3 transition with a polarization 75$\%$ $\sigma$-, 23$\%$ $\pi$ and 2$\%$ $\sigma$+ by intensity  and was left on continuously. The net effect of L1 and L2 was to drive the $^{85}$Rb atoms into the F=2, $m_F=-2$ state. While this all-optical technique was effective in spin-polarizing the atoms, L1 created too much light-assisted collisional loss despite a relatively low intensity. We measured a loss rate of 1.50(75)x10$^{-12}$cm$^3$/sec with 12.1$\mu$W/cm$^2$ L1 intensity. This loss rate was too large for our system for use during 2-CAZ, although we still used this technique for initializing the $^{85}$Rb $m_F$ population distribution. For 2-CAZ cooling, we instead used an alternative spin polarization technique. 

\begin{figure}
\includegraphics{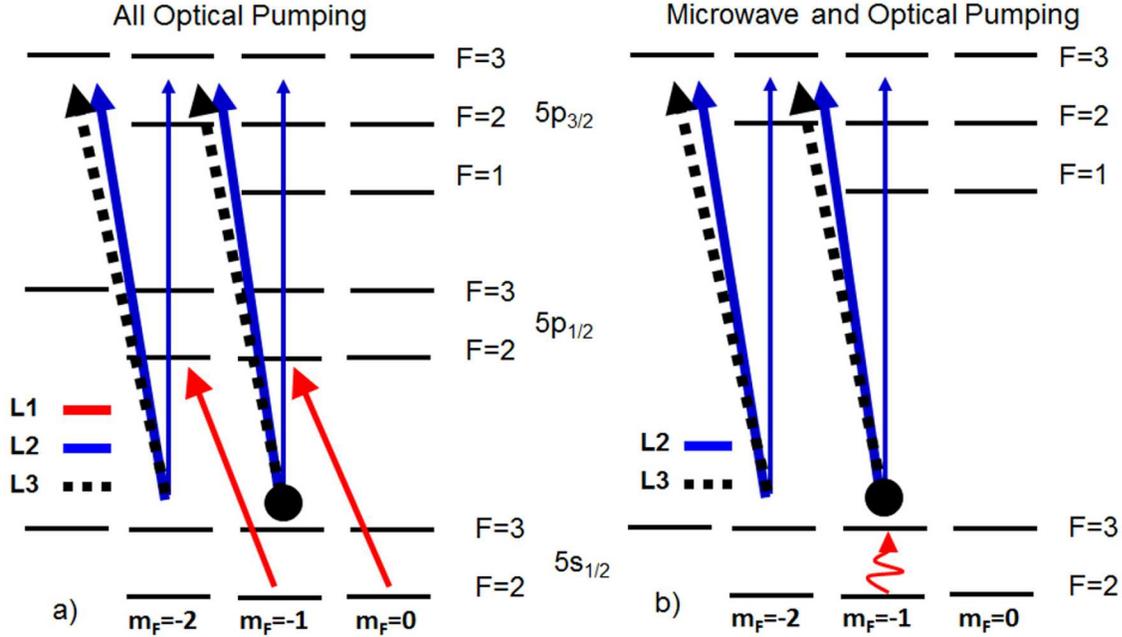}
\caption{Diagram of the laser transitions used for spin-polarization of $^{85}$Rb. All Optical Pumping for Spin-Polarization of $^{85}$Rb is shown in (a.) Direct optical pumping (red/grey) uses circular polarized light from L1 to drive transitions of F=2 atoms. The repump lasers, L2 and L3, are set to have pure $\sigma$- polarization (dashed) and 75$\%$ $\sigma$-, 23$\%$ $\pi$ and 2$\%$ $\sigma$+ polarization (solid black/blue) respectively. The $\sigma$+ component is not shown since it is quite small. Microwave and optical pumping  (b) uses a microwave transition (red/grey) to move atoms into the upper hyperfine ground state. The two repump lasers (those resonant with the F=3 state) are again set to have pure σ- polarization (dashed) and 75$\%$ $\sigma$-, 23$\%$ $\pi$ and 2$\%$ $\sigma$+ polarization (solid black/blue). The goal of each pumping method is to put atoms into a state with a high probability to end up in the F=2, $m_F=-2$ ground state. We have included only the most relevant excited state hyperfine levels. They are not shown to scale. The hyperfine splitting between the ground state, F=2 and F=3 states is 3036MHz, the splitting between the 5P$_{1/2}$ F'=2 and F'=3 states is 362MHz, the splitting between the 5P$_{3/2}$ F'=1 and F'=2 state is 30MHz and between the F'=2 and F'=3 state is 63MHz.}
\end{figure}

\textbf{Microwave and Optical Pumping.}
The microwave pumping scheme was developed as an alternative to the all optical pumping scheme. 
One microwave pumping cycle consisted of a 10ms microwave adiabatic rapid passage sweep spanning 100kHz centered on the transition from F=2, $m_F =-1$ to the F=3, $m_F =-1$ state followed by a 7ms pulse from L2 (figure 5b). Pure $\sigma$- light from L3 is on continuously during the whole sequence. The main advantage of the microwave system over the all optical system is that it avoids the light assisted collisional losses that were produced in the all optical scheme. 

\textbf{RF Scrambler.}
A ten turn coil was located inside of the vacuum chamber and was suspended above the trapping region. Every 80ms an RF sweep was sent to this coil using a function generator. This sweep was centered at 1.4MHz and spans 200 kHz. This frequency was chosen appropriately so that the sweep scrambles the population of the $^{87}$Rb atoms at a 2G field, preventing the accumulation of $^{87}$Rb population in any one of the $m_F$ states.

The most straightforward way to measure the cooling rate would have been to eliminate any other significant heating and/or cooling of the trapped gas, apply 2-CAZ, and measure the resulting gas temperature as a function of time. This straightforward measurement was precluded, however, by unfavorable initial conditions resulting from the interference of the two Rb isotopes during FORT loading, as reported in Ref~\cite{hamilton:2012}. The initial density was not sufficient for a rapid 2-CAZ cooling rate as compared to the background gas-limited 5 second lifetime of the FORT. Further, the gas temperature after initial loading was too large compared to the trap depth to avoid a substantial amount of evaporative cooling.

The presence of this evaporative cooling masked the cooling due to 2-CAZ. Any 2-CAZ cooling largely just reduced the evaporative cooling rate and so the temperature evolution as a function of time changed only modestly (figure 4). In addition to the data, figure 4 also shows the results of a model calculation based on the experimental parameters of the associated measurement. This model calculation was performed by numerically solving a set of first-order differential equations that tracked $^{87}$Rb and $^{85}$Rb $m_F$ state populations as well as the total energy in the gas as a function of time.  Effects such as evaporative cooling, spin-exchange collisions, optical pumping, periodic $^{87}$Rb $m_F$ state scrambling, three-body recombination, and background gas trap loss were included in the model. As shown in figure 4, the model confirmed the insensitivity of the temperature evolution to 2-CAZ cooling in the presence of evaporation.

\begin{figure}
\includegraphics{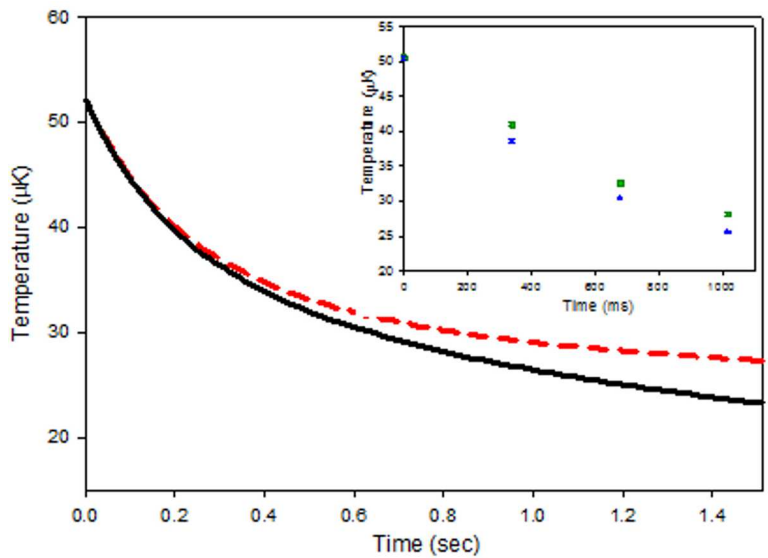}
\caption{2-CAZ cooling in the presence of significant evaporation. The main figure shows predicted temperature vs. time for our experimental conditions with(solid/black) and without(dashed/red) 2-CAZ cooling. The cooling in the no-2-CAZ case is due to evaporation. The inset includes measured no-2-CAZ (square/green) and 2-CAZ (round/blue) experimental results. Our data collection rate mandated that the two data sets be taken on consecutive days. For better comparison a small offset was applied to the 2-CAZ data to overlap the initial starting temperature data points.}
\end{figure}

Given the relative insensitivity of the temperature vs. time evolution to 2-CAZ, we instead used an alternative technique to determine the 2-CAZ cooling rate. The net cooling rate was determined by measuring the steady-state value of the $^{85}$Rb F=1, $m_F=-1$ population right before the microwave sweep during 2-CAZ cooling, combining that population measurement with a measurement of the optical pumping rate, and performing auxiliary experiments to determine the net heating imparted during the optical pumping cycle (i.e. the $\kappa$ parameter in equation (1)).

To characterize the microwave optical pumping transfer efficiency out of the $^{85}$Rb $F=1, m_F=-1$ state we monitored the population in the  $m_F=-1$ state of $^{85}$Rb alone as a function of a chosen number of applied microwave optical pumping cycles. From this data we extracted the on-resonant adiabatic rapid passage transfer probability and the only significant off-resonant adiabatic rapid passage excitation probability of the next-nearest frequency transitions. To obtain these probabilities we initialized the $^{85}$Rb $m_f$ populations using all-optical optical pumping so that the atoms were in the $m_F=-1$ and $m_F=-2$ states. From a rate equation model and the measured L2 and L3 intensity ratio and detunings we calculated that an atom transferred into the F=3 $m_F=-1$ state has a 63$\%$ chance of being driven into the $m_F=-2$ state.  This value does not take into account reabsorption effects, consistent with our observations. Our data yielded an on-resonance microwave excitation probability of 0.76(9) and off-resonance excitation probability of 0.0047(8). We independently confirmed the on-resonance excitation probability by exciting all the $m_F$ states in turn and looking at the average excitation probability.  

We then measured the steady-state (i.e.\ after many 2-CAZ cooling cycles) population in the $m_F=-1$ state via microwave sweep during 2-CAZ cooling. To ensure the system had reached steady-state, this measurement was performed after 30 microwave/optical pumping cooling cycles of 17ms duration. At the time this measurement was performed the full $^{85}$Rb number was 1.54(6)x10$^6$ with a density of 4.8(2)x10$^{11}$cm$^{-3}$ and the full $^{87}$Rb number was 1.11(6)x10$^6$ with a density of 3.4(2) x10$^{11}$cm$^{-3}$, where all of these quoted uncertainties reflect only statistical uncertainty. Systematic uncertainties in the number and density are about 20$\%$ for our system. The temperature of the atoms was 36.8(5)$\mu$K. The results of this measurement indicated a steady state fraction of $^{85}$Rb atoms in the $m_F=-1$ state of 0.090(12) for these conditions.

The cooling rate in the absence of any heating mechanisms can be determined from the optical pumping efficiency and the steady-state $m_F=-1$ population since the combination of these two quantities can be used to calculate the Zeeman energy extracted from the gas in an optical pumping cycle. We term this the maximum cooling rate. The results described above imply a maximum cooling rate of 10.8(9)$\mu$K/sec given the applied magnetic field of 2.0 G. This can be compared to the prediction of equation (1) with $\kappa$ set to be 0. In order to avoid systematic uncertainties in the determination of the density, we use an experimentally determined value for the weighted spin exchange rate, which is consistent with theoretical predictions~\cite{Kokoouline}. The corresponding maximum cooling rate from equation (1) is 11.6$\pm$1.9 $\mu$K/sec. Thus, the measured maximum cooling rate and equation (1) are in agreement.

After determining this maximum cooling rate, we proceeded to account for heating during the optical pumping cycle to determine a net cooling rate. To measure the heat and loss associated with the microwave optical pumping technique we performed an auxiliary experiment. 12 microwave optical pumping cycles were applied to a gas of $^{85}$Rb alone. To ensure the atoms were optically pumped in a closed cycle, we changed the microwave frequency to be resonant with the degenerate $^{85}$Rb F=2 $m_F=-1$ to F=3 $m_F=-2$/F=2 $m_F=-2$ to F=3 $m_F=-1$ transitions. These 12 cycles increased the gas temperature by 4.89(56)$\mu$K. During these 12 cycles we also lost 16(3)$\%$ of our atoms, which we experimentally observed to be density-dependent but not optical-pumping-light-dependent. The observed loss was consistent with loss expected from hyperfine-changing collisions for our atom density ~\cite{Burke1998,Burke1999}. By assuming that part of the 4.89(56) $\mu$K temperature increase was due to two-body collisions in a harmonic trapping potential ~\cite{twobody}, we determined the density-dependent portion of the heating to be 2.53(57)$\mu$K. By converting between the density and temperature conditions for this auxiliary experiment and our 2-CAZ experimental parameters we determine the density-dependent heating rate relevant to the 2-CAZ cooling measurement to be 1.35(7)$\mu$K/sec.

The remaining 2.35$\mu$K of heating observed in the auxiliary experiment is owing to random momentum recoils from the photon scattering required for optical pumping. This recoil heating was consistent with the calculated number of photons required in the optical pumping cycle after taking into account the direction of the optical pumping beam. A redesign of our apparatus to align L3 to be exactly perpendicular (rather than 45$^\circ$) to the optical trap axis would reduce this heating by a predicted factor of 2. This photon-scattering-induced heating rate was converted from the auxiliary experiment to the relevant rate for 2-CAZ cooling by adjusting for the fraction of $^{85}$Rb atoms that cycle through the F=3 state to be 1.48(8)$\mu$K/sec.

We did not observe any statistically significant heating due to off-resonant light-assisted collisions caused by the optical pumping light, consistent with our expectations. We do expect heating during the optical pumping cycle due to $^{85}$Rb atoms in F=3 state undergoing spin-exchange collisions. Since the $^{85}$Rb F=3 state has a g$_F$ factor opposite in sign from the F=2 state these collisions result in a energy increase of 5/6$\mu_B$B instead of the expected decrease of 1/6$\mu_B$B from an F=2 collisions. The loss attributed to hyperfine-changing collisions described above prevented us from measuring this rate directly, but it can be esimated to be 0.25(7)$\mu$K/sec for our conditions.

After subtracting these heating rates from our maximum cooling rate we find a net cooling rate of 7.72(91)$\mu$K/sec. It is clear that the heating rates have reduced our cooling capacity significantly even at a gas temperature much greater than the recoil energy. We can convert the measured heating rates determined above to a value for $\kappa$ through using the cycle length, optical pumping efficiency, and relative number of $^{85}$Rb and $^{87}$Rb atoms such that $\kappa/k_B$=6.36(59)$\mu$K. Using this value, equation (1) predicts a net cooling of 8.1$\pm$1.4 $\mu$K/sec, again in agreement.

To help insure that we did not miss a significant heating or cooling factor, we performed an additional experiment. We loaded the FORT with $^{85}$Rb and $^{87}$Rb at a reduced trap depth and held the atoms there for 1 second before adiabatically ramping up the FORT to its full depth. Following this sequence radically reduced the evaporative cooling rate at the expense of a factor of $\sim$5 in the number of the atoms in the FORT. We then applied 60 17ms pulse 2-CAZ cycles. At the end of the 2-CAZ cycle there were 0.24(1)x10$^6$ $^{85}$Rb atoms and 0.21(1)x10$^6$ $^{87}$Rb atoms. While not optimum for a 2-CAZ cooling rate, these conditions were sufficient to compare a measured amount of 2-CAZ-related cooling to our differential equation model predictions (c.f.\ fig. 4). We measured a temperature reduction of 1.92(27)$\mu$K as compared to a model prediction of 1.98(50)$\mu$K. This provides additional evidence that we have not left out any substantial heating or cooling considerations.

\section{Discussion of results and implications for the utility of 2-CAZ}

Initially, the goal of this work was to use 2-CAZ for efficient cooling of an $^{85/87}$Rb mixture to few $\mu$K temperatures. This was not achieved in part because of the difficulty in creating sufficient initial conditions, particularly sufficiently high initial density, due to difficulties in simultaneous $^{85/87}$Rb optical trap loading~\cite{hamilton:2012}. In principle this limitation can be overcome though the use of different trapping techniques such as hybrid magnetic/optical trap loading, pre-cooling, or trap geometry modification to increase initial density. However, our measurements reported above indicate that sufficient initial densities will not be the only critical factor in an $^{85/87}$Rb mixture for efficient 2-CAZ cooling. This in turn has implications for evaluating when 2-CAZ may be useful in other systems with other types of atoms.

Once $k_B T$ has been lowered to be on the order of $\kappa$, the cooling rate will start to decrease exponentially (c.f.\ Eq. (2)). Thus, $\kappa$ is an indication of the lowest practical achievable temperature. Ideally, $\kappa$ would be on the order of a few photon recoil energies ($\frac{(\hbar k)^2}{2m}=k_B \cdot 0.19\mu$K for Rb). The measured value of $\kappa$ presented above is far above this ideal case.  There are three main contributing factors. First, more efficient optical pumping would reduce $\kappa$ by about a factor of 2 overall. Second,the heating contribution from photon recoils for the actual implementation of our optical pumping is much higher than theoretically possible. Finally, the $^{85}$Rb atoms are spending ``too long'' in the F=3 state resulting in significant heating from hyperfine-changing collisions. The second problem can be improved though multiple, although non-trivial, modifications to our apparatus. The final problem likely poses a severe challenge for the utility of 2-CAZ for $^{85/87}$Rb as it is density-dependent. For instance, if the cooling were made to be more effective and the densities increased by an order of magnitude over our conditions reported above, an order of magnitude reduction in time the $^{85}$Rb atoms spend in the F=3 state during optical pumping would be required just to maintain the current $\kappa$ contribution from hyperfine-changing collisions, let alone reduce it.

The time the $^{85}$Rb atoms spend in the F=3 state is currently limited by the microwave power, but the timescales associated with L2 and L3 are less that an order of magnitude away from this microwave limit. While there are no reabsorption effects observed for the reported conditions above, we do observe degradation in optical pumping efficiency at an order-of-magnitude more intensity for L2 and L3. Thus, the window of sufficiently high density for effective cooling and sufficiently low time spent in the F=3 state appears to be small.

A potential solution to this problem would be to perform 2-CAZ by optically pumping an atom that had no hyperfine structure but a non-zero electron spin ground state (e.g.\ $^{52}$Cr). This avoids the issues associated with the hyperfine changing collisions and would likely lead to better performance as long as the atom used did not suffer the unusually high light-assisted collision rate observed for low intensities in our implementation of the all-optical optical pumping of $^{85}$Rb. In order for 2-CAZ to be effective, the other atom or molecule in such a cooling situation would have to have hyperfine structure to produce an energy barrier ($\Delta$) for spin-exchange collisions.

Despite the limitations that we observed in 2-CAZ for $^{85/87}$Rb, there are still reasons to investigate this cooling in a more favorable system. In many light-based non-evaporative cooling techniques, the lowest achievable temperatures are limited by the density of the atoms being cooled through effects like reabsorption. Having the freedom to adjust the density of the optically-pumped isotope while still maintaining a high density of the non-optically-pumped isotope can result in improvements in the achievable cooling rate at low temperatures. For 2-CAZ cooling, this can be seen by assuming that the cooling is taking place under conditions where $\kappa$ is proportional to the number of optically-pumped atoms $N_2$, i.e.\ $\kappa=\alpha k_B N_2$, where $\alpha$ is constant. For this $\kappa$, in the limit of fast optical pumping and $N_2 << N_1$ the optimal cooling rate is 

\begin{equation}
\frac{dT}{dt}=-\frac{k_2N_2}{V}\frac{T}{3}exp(-1-\frac{\alpha N_2}{T})
\end{equation}

\noindent where $V$ is the effective volume of the gas such that the average density of the type 2 atoms is $\frac{N_2}{V}$. For $\alpha N_2>T$, reducing $N_2$ increases the cooling rate.  Compared to cooling in a gas with only one type of atom present, this ability to increase the cooling rate by reducing $N_2$ under reabsorption-limited conditions can produce orders-of-magnitude improvements in the net cooling rate. This is a general feature of non-evaporative cooling with two different types of atoms instead of one and is not limited to 2-CAZ. The predicted improvement in low-temperature cooling capabilities is an intriguing area of investigation if a system better-suited to 2-CAZ cooling than $^{85/87}$Rb can be used.

In summary, we have extended CAZ cooling from a single-atom cooling case to one in which two different types of atoms are trapped and cooled. We have discussed the advantages of using two different types of atoms as compared to one for CAZ cooling. We experimentally measured 2-CAZ cooling in an $^{85/87}$Rb mixture and found that the measured cooling rate agrees with a simple analytical expression for the predicted cooling rate. This allowed us to interpret the limitations of 2-CAZ cooling for our experimental configuration. Cross-isotope interference in optical trap loading and the hyperfine-changing collision rate in the $^{85/87}$Rb mixture prevented the observation of robust 2-CAZ cooling in our measurements and would be challenging to overcome.  Our analysis indicates the requirements that would be necessary for a better mixture of atoms for 2-CAZ cooling. Predicted cooling rate improvements at low gas temperatures with the use of two different types of atoms remains an intriguing possibility for 2-CAZ cooling.

\ack
This work is funded by the Air Force Office of Scientific Research, Grant No. FA9550-08-1-0031.

\bibliographystyle{unsrt_mod_iop}

\end{document}